\documentstyle[prl,aps,twocolumn,epsf]{revtex}
\def\break#1{\pagebreak \vspace*{#1}}
\begin{document}

\draft

\title{On Wolter's vortex in total reflection}

\author{H.C. Rosu$^{a,b}$ %\cite{byline} %
%$^{\dagger}$
%$^{\ddagger}$
}

\address{ $^{a}$
{Instituto de F\'{\i}sica de la Universidad de Guanajuato, Apdo Postal
E-143, Le\'on, Gto, M\'exico}\\
$^{b}$
{Institute of Gravitation and Space Sciences, Bucharest, Romania}\\
}

%\date{submitted to X, mmddyy}
\maketitle
\widetext

\begin{abstract}

ABSTRACT. The `circulatory wave' (``die zirkulierende Welle") put into
evidence in 1949 by Wolter (Wolter's vortex) in total reflection is
interpreted as a phase defect in the scalar theory of Green and Wolf
of 1953, which is the Madelung (hydrodynamic) representation of the optical
field.
Some comments are added on its possible relevance for the Hamamatsu
experiment aimed to clarify the wave-particle duality at the ``single photon"
level of down-converted laser beams.

\end{abstract}

\vskip 0.1cm

RESUMEN. La `onda circulante' descubierta en 1949 por Wolter
(vortice de Wolter) en la reflexi\'on total, es interpretada como un
defecto de fase en la teor\'{\i}a escalar de Green y Wolf de 1953, la cual
es la representaci\'on de Madelung del campo \'optico. Algunos
comentarios son vertidos sobre su posible relevancia en el experimento
Hamamatsu intentado para clarificar la dualidad onda-part\'{\i}cula al nivel
de un solo fot\'on obtenido de rayos laser en conversi\'on descendente.

\pacs{PACS numbers:  42.10.Fa; 03.65.-w   \hfill
Rev. Mex. Fis. 43 (1997) 240-247
%LAA number: quant-ph/9506015
}

%%%%%%%%%%%%%%%%%%%%%%%%%%%%%  THE PAPER  %%%%%%%%%%%%%%%%%%%%%%%%%%%%%%%%%%
%%%%%%%%%%%%%%%%%%%%%%%%% written by H.C. Rosu  %%%%%%%%%%%%%%%%%%%%%%%%%%%%
%%%%   1979-1980 in Bucharest, 1992 in Trieste, 1995-1996 in Leon/Gto  %%%%%
%%%%%%%%%%%%  Revista Mexicana de Fisica 43 (1997) 240-247   %%%%%%%%%%%%%%%
%%%%%%%%%%%%%%%%%%%%%%%%%%%%%%%%%%%%%%%%%%%%%%%%%%%%%%%%%%%%%%%%%%%%%%%%%%%%
\narrowtext

%PACS: 42.10.Fa - Edge and boundary effects, refraction

%PACS: 03.65.-w - Quantum theory; quantum mechanics

%%%%%%%%%%%%%%%%%%%%%%%%%%%%%%%%%%%%%%%%%%%%%%%%%%%%%%%%%%%%%%%%%%%%%%%%
{\bf 1. Introduction}\\
At the present time optical phase defects are quite well known especially
due to their connection with transverse laser patterns \cite{lp}.
Less well known
is an old phase singularity discovered by Wolter \cite{w} in
total reflection (TR). Even though Wolter's vortex plays an important
role in the realm of TR, it has not drawn much attention over the years.
Checking the literature one will find out that this vortex
has not been discussed since 1970, in the thesis of Lotsch \cite{l}.
It is true however
that very similar waves can be encountered in other research fields, e.g.,
in Madelung's quantum mechanics where more detailed analyses have been made
\cite{h}.

TR is one of the experimental discoveries of Newton,
so it has a long history of 300 years \cite{l,s} and a long
list of authors. For a recent review, though on the frustrated version,
 see \cite{z}.

This phenomenon (or more exactly frustrated internal TR) is
similar to the tunnel effect in quantum mechanics, an analogy presented
in many textbooks \cite{ber}. This is so, because, despite
the name, there is some
light in the rarer medium even at the limiting angle of total
reflection and beyond it. Fresnel called this faint field
{\em evanescent light}.

The evanescent light is not really a wave, not even a wave packet but
only part of a wave packet. Thus, in TR experiments it is possible to
 reveal some remarkable inhomogeneities of wave
%\break{1.8 in}
 packets which escape observation in the usual propagation.

On the experimental side, only in 1947 were Goos and H\"{a}nchen
\cite{gh}
%\break{1.95 in}
able to observe the longitudinal shift of the totally reflected
light calculated by Picht \cite{p} already in 1929. Another shift,
%\break{1.97 in}
a transverse one, has been predicted (with some prehistory)
by Fedorov \cite{f} in 1955
%\break{2.01 in}
and discussed either in the Poynting-vector method or the stationary-phase
\break{2.15in}
one by a number of authors. It was first investigated
experimentally by Imbert \cite{i} in 1969. Very recently,
Dutriaux, Le Floch, and Bretenaker \cite{dfb} made important progress by
measuring the transverse
displacement of the helicoidal eigenstates of a He-Ne laser beam for
various angles of
incidence, providing in this way a first check of the various
formulas for the transverse shift for circular polarization.

A simple understanding of the longitudinal shift may be obtained in terms of
the propagation of light in the second rarer medium very close
to the surface in such a way (Newton's parabola) that apparently the incident
ray is
reflected by a surface located at some small depth in the second
medium. This depth gives the exponential rate of decrease of the field
amplitude.
For the transverse shift, Costa de Beauregard \cite{cb}
has given a quite
unconventional explanation, which he dubbed the ``inertial spin effect",
essentially the noncolinearity of the velocity and momentum of the
photon in TR conditions. The problem of the angular and spin momentum
conservation at reflection is still
debatable \cite{pl}. Both shifts are polarization dependent and
their order of
magnitude is about half a wavelength in the case of the
transverse shift and usually ten wavelengths for the longitudinal one.

The organization of the paper is as follows. Section 2 is a brief
summary of
Wolter's results on the `circulatory wave'. In the next
section, after an outline of the Green-Wolf (GW) scalar theory,
Wolter's vortex is explained within the GW theory as a result of the
`quantization' of
the `velocity' circulation.
The next section contains some comments on the possible relevance of
Wolter's vortex in the Hamamatsu wave-particle experiment with laser beams at
the single-photon level, and I end up with concluding remarks.

\bigskip
%\vspace*{1cm}

{\bf 2. The Circulatory Wave}

The circulatory wave was discovered when Wolter \cite{w}
performed accurate measurements of the Goos-Hanchen shift
by means of his powerful method called ``minimum ray characteristic".
Wolter observed the displacement of interference {\em minima} formed by
{\em two} slightly inclined (plane) waves (and the two reflected ones) which
have undergone multiple reflections,
on one hand at the interface glass-air, and on the other hand at the
interface glass-silver. A plane parallel plate was used on which a
silver strip of $\lambda /4$ thickness was deposited. The advantages
of this technique over the diffraction maximum method were presented
by Wolter himself \cite{w1}.
The ``minimum ray characteristic" allowed Wolter to reveal clearly
the details of the electromagnetic field in the immediate neighborhood
of the interface between the two media. He has also calculated
the phase surfaces, energy flux lines and time averaged Poynting
vectors, obtaining a full picture of the field. Wolter was
 particularly careful with the region close to the geometric point of
 TR (denoted as the origin in his paper), where he put into evidence the
 circulatory wave of the energy streamlines,
 both experimentally and in theory, as a special
 solution that comes out when integrating a Bernoulli differential equation
 of order $n=-1$, i.e., of the
 type $ y^{'}= a(z)y^{-1} + b(z) y $,
with $y^{'} =\frac{dy}{dz}$, $a(z)=z(1+2\pi f z)$ and $b(z)=1/F=$ const.,
in almost
Wolter's notations; $y$, which is the coordinate along the surface in the
thin medium,
the transverse coordinate $z$, and $f\approx 2\pi F$ are related to the
mean angle of incidence (angle of observation of the interference minima) and
the mean shift.
%%%%%%%%%%%%%%%%%%%%%%%%%%
%The general solution of the Bernoulli eq. reads
%$$
%y^2=\exp(2\int_z_{0}^{z}b(z')dz')(a_1+2\int_z_{0}^{z}a(z')\exp(2\int_z_{0}^z
%b(z')dz')
%$$
%leading to
%$$
%y^2=\exp(2(z-z_0)/F)(a_1+\frac{2}{F}\int_z_0^z z(z-z_0)(1+2\pi fz)dz)
%$$
%[E.A. Bartnik et al., PL A {\bf 204}, 263 (1995)] [June 12/96]
%%%%%%%%%%%%%%%%%%%%%%%%%%%%%%%%

The circulatory
 wave is very similar to an ordinary vortex in the turbulent
 motion of a classical fluid. In the rarer medium there is always
 a saddle point (or a stagnation point in hydrodynamical language),
 which is essential for the energy flow in TR. At that point the
 incoming wave is broken in two parts: one is going through the glass
 and after one period is again crossing the surface in order to meet the
 other part which remained in air. The splitting and meeting are
  repeated every period. Therefore the permanence of the evanescent
wave in the rarer medium is due to the circulatory wave, which is carrying
energy back and forth across the interface.
  We have here a clear explanation for the fact that although the
  incident energy is equal to the reflected one, there is also a time
averaged Poynting vector parallel to the surface propagating very
close to the surface ($d\approx \lambda$) in the rarer medium. Before Wolter,
the energy problem in TR has been considered by many authors, e.g., by
Drude \cite{d} for the
plane wave case, and by Picht \cite{p}, Noether \cite{n},
and Schaefer and Pich \cite{sp} for more realistic situations.
  These authors have shown that due to the limited extent of the beam field,
some amount of energy is extracted from
one side of the incident beam and added to the other side of the
reflected beam, in the same plane of incidence, after having
  propagated parallel to the interface in the less dense medium. At the
  saddle point the energy flux is zero. Between the stagnation point
  and the interface, in the region near the origin, we enter
  the vortex rings of the circulatory wave. Most of the core of the
   vortex wave is to be found in the denser medium and in the middle
    of the vortex
 the amplitude of the wave is naught, but in the rarer medium the
 amplitude never goes to zero on the scale of a wavelength.

Wolter obtained the circulatory wave by the interference of four plane
waves (two incident and two reflected ones). Vortex configurations never
show up in the TR of a single plane wave. Braunbek \cite{b} proved that the
 field of only three plane waves displays vortex configurations.

\bigskip
%\vspace*{1cm}

{\bf 3. Wolter's vortex in the GW scalar theory}

The GW scalar theory of electromagnetic
field has been developed in the 1950's \cite{gw}, and is nothing but the
electromagnetic counterpart of the Madelung quantum theory. The GW
theory is based on a single, generally complex scalar wave function
{\em V({\bf x},t}) and is
valid in regions free of charges and currents. This theory was
extended to the full generality by Roman \cite{r}.

The transition to the GW scalar theory is especially simple in the case of the
electromagnetic field in vacuum, that can be completely specified by the
vector potential, {\bf A}({\bf x},{\it t}) satisfying the
divergence free condition
%%%%%%%%%%%%%%
$$div\;{\bf A}({\bf x},{\it t})=0~.   \eqno(3.1)$$
%%%%%%%%%%%%%%%
In Fourier integral form {\bf A} reads
%%%%%%%%%%%%%%
$${\bf A}({\bf x},{\it t})= \int [{\bf a}({\bf k},t)\cos({\bf k,x}) +
{\bf b}({\bf k},t)\sin({\bf k,x})]d{\bf k}~. \eqno(3.2)$$
%%%%%%%%%%%%%%
The integration is taken over the half plane $k_{z}>0$. Eq.~(3.1) implies
that for each {\bf k} we have
%%%%%%%%%%%%%%%%%%%%%%%
$${\bf k\cdot a}={\bf k\cdot b} = 0~.    \eqno(3.3)$$
%%%%%%%%%%%%%%%%%%%%%%%
One may introduce two real, mutually orthogonal unit vectors
 ${\bf l}_{1}({\bf k})$ and ${\bf l}_{2}({\bf k})$, both at right angles
 to ${\bf k}$
 %%%%%%%%%%%%%%%%%%
 $${\bf l}_{1}({\bf k})=\frac{{\bf n \times k}}{|{\bf n \times k}|}
      \eqno(3.4)  $$
%%%%%%%%%%%%%%%%%%%%
 $${\bf l}_{2}({\bf k})=
 \frac{{\bf k} \times {\bf l_{1}}}{|{\bf k} \times{\bf l_{1}}|}~,
   \eqno(3.5) $$
%%%%%%%%%%%%%%%%%%%%
where {\bf n} is a real, arbitrary but fixed vector. The two vectors
 ${\bf a}$ and ${\bf b}$ may be expressed in the form
 ${\bf a}({\bf k},{\it t})=a_{1}{\bf l_{1}} + a_{2}{\bf l_{2}}$ and
${\bf b}({\bf k},{\it t})= b_{1}{\bf l}_{1}+ b_{2}{\bf l}_{2}$.
One may pass to the complex combinations
$\alpha = a_{1}+ia_{2}$ and $\beta=b_{1}+ib_{2}$, which
are considered as the Fourier
coefficients of a new function, known as the complex potential of the field
%%%%%%%%%%%%%%%%%%%%
 $$V({\bf x},{\it t})=\int[\alpha({\bf k},{\it t})\cos({\bf x},{\it t})+
 \beta({\bf k},{\it t})\sin({\bf x},{\it t})]d{\bf k}~.   \eqno(3.6)  $$
 %%%%%%%%%%%%%%%%%%%
 Once the constant vector ${\bf n}$ has been chosen, the complex
  potential {\it V} is uniquely specified by the Fourier components
  ${\bf a}$ and ${\bf b}$ of the vector potential ${\bf A}$ and hence by
  ${\bf A}$ itself. In this way the vector potential ${\bf A}$ is
  uniquely specified by the complex potential ${\it V}$.
Green and Wolf have shown that the relationship between ${\bf A}$ and
${\it V}$ is a linear one
%%%%%%%%%%%%%%%%%%%
$$ {\it V}({\bf x},{\it t})= \int {\bf A}({\bf y},{\it t})\cdot
{\bf M}(y-x)dy~.    \eqno(3.7)  $$
%%%%%%%%%%%%%%%%%%%%%
The kernel ${\bf M}$ is the Fourier transform of the set of the complex
base vectors ${\bf L}({\bf k})$ given as follows
%\break{0.03 in}
%%%%%%%%%%%%%%%%%%%%
$${\bf L}({\bf k})= {\bf l_{1}}({\bf k})+ i{\bf l_{2}}({\bf k}),
\; k_{z}>0~.   \eqno(3.8)  $$
%%%%%%%%%%%%%%%%%%%%%%%%%%

The form of the momentum
density ${\bf g}({\bf x},{\it t})$ and the energy density
 $w({\bf x},{\it t})$, in terms of the complex
  potential are similar to the probability current and probability
  density in quantum mechanics
%%%%%%%%%%%%%%%%%%%%%%%
$${\bf g}({\bf x},{\it t})=-\frac{1}{8\pi}[{\dot V}^{*} \nabla V +\dot{V}
 \nabla{ V^{*}}]~,      \eqno(3.9)  $$
 $$w({\bf x},{\it t})=\frac{1}{8 \pi}[VV^{*}+\nabla V \nabla V^{*}]~,
   \eqno(3.10)  $$
%%%%%%%%%%%%%%%%%%%%%%%%
where the dot denotes the time derivative and the star is the complex
conjugate operation. The close connection to the quantum potential
representation of quantum mechanics is natural if we think that both
frameworks are hydrodynamical-like theories.

The energy density and the energy flow vector satisfy an equation of
continuity of the usual form $\frac{\partial w}{\partial t}+div\; {\bf g}=0$.

If the complex potential is written down in the polar form
$ V({\bf x},{\it t})=A({\bf x})\exp[iB({\bf x},{\it t})]$,
from Eq.~(3.9) one gets
%%%%%%%%%%%%%%%%%%%%%%
$${\bf g}({\bf x},{\it t})=-\frac{1}{4\pi}(\dot{A} \nabla A + A^{2}
 \dot{B} \nabla B)   \eqno(3.11)$$
%%%%%%%%%%%%%%%%%%%%%%
and by making use of the eikonal phase ansatz $B({\bf x},{\it t})=
kS({\bf x})-\omega t $, one will find
%%%%%%%%%%%%%%%%%%%%%
$$ {\bf g}({\bf x},{\it t})=\frac{k^{2}}{4\pi}A^{2}\nabla S~. \eqno(3.12)$$
%%%%%%%%%%%%%%%%%%%%
Thus, there is an orthogonal flow of the energy onto
the surfaces $S({\bf x})=$ const,
which are the wavefronts. The momentum density may be
looked upon as curves (rays) orthogonal to wavefronts. By means of the
hydrodynamical definition of the velocity field ${\bf v}=\nabla S $,
one can find out the irrotational condition $ \nabla \times {\bf v} = 0 $
everywhere except at the nodal regions $A=0$, where S cannot be defined.

Introducing in the usual manner the circulation as the line integral
$ \oint_{L}{\bf v}d{\bf r}$
on a closed path $L$, we obtain the well-known condition of its
discrete valuedness provided the loop is not passing directly through
 any of the nodal regions of the scalar potential. The quantization of
the circulation
%%%%%%%%%%%%%%%%%%%%%%%
$$ \oint_{L} {\bf v}d{\bf r} =2\pi n \;,\; n= 0,1,2,....~,  \eqno(3.13)  $$
%%%%%%%%%%%%%%%%%%%%%%
is an old hydrodynamical fact due to the continuity and single-valuedness
character of the velocity potential.
More general circulations may be found in superconductivity and
superfluidity as well as in various types of quantum field models.
In the framework of the simple scalar (hydrodynamical) theory, the
discrete values of the velocity circulation expresses the fact that
the phase $S({\bf x})$ of the scalar potential is defined up to an
integer multiple of $2\pi$.

If within the loop there is only one node of the scalar potential,
as in the case of Wolter's wave, the circulation will be
 $2\pi$ corresponding exactly to a time delay of one period in the
 denser medium.
The nodal regions of the scalar
potential occur because of the diffractive conditions at the surface.

%%%%%%%%%%%%%%%%%%%%%%%%%%%%%%%%%%%%%%%
There are obvious similarities between the GW theory and
the known superfluids (superconducting electrons,
superfluid He$^{4}$, and superfluid He$^{3}$),
 as we already remarked. The macroscopic quantum-mechanical
wave function of superfluids should be single valued. This
implies the quantization of circulation in a neutral superfluid and of
magnetic flux in a charged superfluid. A complete microscopic theory
of superfluid He$^{4}$ does not exist, although the standard
 interpretation is in terms of Bose-Einstein condensation and the
 postulate of quantized circulation in units of $h/m_{4}$.
Some phases of liquid He$^{3}$ are considered to be Cooper-paired,
neutral superfluids with the circulation quantized in units
of $h/m_{3}$. Recently, in \cite{d91} the quantized
  circulation
of the B-phase of He$^{3}$  has been observed in the laboratory. On the other
hand, Wolter's vortex is equivalent to a quantized circulation, however
occuring in a rather special kind of electromagnetic (zero-mass)
superfluid \cite{sf}.
%%%%%%%%%%%%%%%%%%%%%%%%%%%%%%%%%%%%%%%%%
\bigskip
%\vspace*{.7cm}

{\bf 4. Wolter's vortex and Hamamatsu experiment}

Wolter's vortex in TR might be relevant for the
progress in the field of wave-particle duality in the case of light,
especially for the Hamamatsu anticoincidence experiment of
Mizobuchi and Ohtak\'e \cite{mo} in
which both behaviors are revealed simultaneously at `single-photon'
levels of the tested beam as suggested by Ghose, Home, and Agarwal \cite{gha}.
The experiment can be considered as a modern version of the double-prism
experiment on frustrated TR first done by Bose in 1897, and in fact
a Bose's double-prism is used as the beam splitter.
The physical aspect to be understood is the nonlocality
of the photon field in this particular type of experiment. In our opinion, for
anticoincidence experiments performed with Bose beam splitters, Wolter's
vortex is
of direct relevance to the non-locality responsable for the {\em and} logic
of the wave-particle duality. This is so because the Bose beam splitter is
basically operating on the principle of the optical tunneling
(frustrated TR) in which Wolter's vortex is part of the phenomenology.
Moreover, one should be aware of the network of vortices existing within any
laser beam \cite{freund}.
Interestingly, in a paper of Elitzur, Popescu and Rohrlich \cite{epr}
the nonlocality is shown to be present in each pair of an
ensemble of spins when the spins are coupled in singlet states only.
They comment on a formal analogy between local and
nonlocal photons and normal and superfluid components of helium ll.

Other types of measurements at the `single-photon' level
emphasize the non-local nature
of a `single photon' and the relationship with the EPR paradox and Bell's
inequalities, for a review see \cite{hs}.

It would be interesting to perform experiments dealing with `single photons'
in the sense
of one-photon Fock (number) state \cite{hm} in order to investigate the
nonlocal properties of the `single-photon' Fock field.

\bigskip
%\vspace*{1cm}

{\bf 5. Concluding remarks}

In this paper I focused on Wolter's vortex, a forgotten phase
singularity of the optical field in TR, clearly put into experimental
evidence already in 1949.

The vorticity-type singularities are special details of all sorts of
wave packets: electromagnetic, acoustic and/or quantum mechanical ones
\cite{h}. They can be created
at the moment of formation of the wave packet or by diffraction during
propagation. Their space scale is of the order of the wavelength.

We also gave some hints on possible implications of the Wolter's vortex in
the Hamamatsu experiment on wave-particle duality claiming an {\em and} logic
for this duality at the `single-photon' level of the tested beam. It might be
possible that in the case of
Bose (double-prism) splitter, the nonlocality of the `single photon' and
the Wolter's vortex are interrelated. Moreover, the Hamamatsu
experiment must be repeted in more definite conditions and making use of
the recent technological progress in high-efficiency `single-photon' detectors.
A detailed treatment of low-intensity photon beams is needed.

It should also be emphasized that vortical singularities, being classical
interference patterns endowed with a circulation constraint, are not included
in quantization procedures for the evanescent waves \cite{cm}. Very
recently, Lugiato and Grynberg elaborated on the effect of quantum noise on
the optical vortices \cite{lg}.

\bigskip
%\vspace*{1cm}

{\bf Acknowledgment}

This work was partially supported by the CONACyT Project 4868-E9406.

%\vspace*{1cm}

%\vskip 1cm

%\section*{\bf Figure Caption}

%Fig.~1. A vortex configuration with the vortex point at (1,1) and the
%stagnation point at (2,1), which is the MATHEMATICA version of the vortex
%in Fig.~8 of Hirschfelder, Christoph, and Palke, J. Chem. Phys. {\bf 61},
%5435 (1974).

%\vskip 0.5cm

%MATHEMATICA PLOT:
%\vskip 0.2cm

%ContourPlot[(x-1)$^2$+ (y-1)$^2$-2/3*(-y+1)$^3$-1, {x,-.5,3}, {y,-1,2},
%Contour Shading-->False, Contour Smoothing-->Automatic, PlotPoints-->20,
%Contours-->50]

\end{document}